\documentclass[intlimits,twoside,a4paper]{article}

\usepackage{amsmath,amssymb}
\usepackage{graphicx}
\usepackage{wrapfig}

\usepackage[T2A]{fontenc}
\usepackage[cp1251]{inputenc}
\usepackage{cmpj2}

\issue{2012}{15}{1}{13001}

\doinumber{10.5488/CMP.15.13001}

\title[Motifs and entropy of statistically interacting particles]%
{Interlinking motifs and entropy landscapes of statistically interacting particles}

\author[P. Lu \textsl{et al.}]{P. Lu\refaddr{label1}, D. Liu\refaddr{label1},
        G. M{\"u}ller\refaddr{label1}, M. Karbach\refaddr{label2}}

\authorcopyright{P. Lu, D. Liu, G. M{\"u}ller, M. Karbach, 2012}

\addresses{
\addr{label1} Department of Physics, University of Rhode Island, Kingston RI
02881, USA
\addr{label2} Fachbereich Physik, Bergische Universit{\"a}t Wuppertal, 42097
Wuppertal, Germany}

\date{Received November 28, 2011}
\begin{document}

\maketitle

\begin{abstract}
  The $s=1/2$ Ising chain with uniform nearest-neighbor and next-nearest-neighbor coupling
  is used to construct a system of floating particles characterized by motifs of up to six consecutive
  local spins.
  The spin couplings cause the assembly of particles which, in turn, remain free of interaction energies
  even at high density.
  All microstates are configurations of particles from one of three different sets, excited
  from pseudo-vacua associated with ground states of periodicities one, two, and four.
  The motifs of particles and elements of pseudo-vacuum interlink in two shared site variables.
  The statistical interaction between particles is encoded in a generalized Pauli principle,
  describing how the placement of one particle modifies the options for placing further particles.
  In the statistical mechanical analysis arbitrary energies can be assigned to all particle species.
  The entropy is a function of the particle populations.
  The statistical interaction specifications are transparently built into that expression.
  The energies and structures of the particles alone govern the ordering
  at low temperature.
  Under special circumstances the particles can be replaced by more
  fundamental particles with shorter motifs that interlink in only one shared site variable.
  Structures emerge from interactions on two levels: particles with shapes from coupled spins
  and long-range ordering tendencies from statistically interacting particles with shapes.
\pacs 05.50.+q, 75.10.-b
\end{abstract}

%
\section{Introduction}\label{sec:intro}
%
Condensed matter is an aggregate of interacting particles.
The interactions operate in hierarchies regarding strength and range.
Atomic nuclei are composed of strongly coupled protons and neutrons.
Electrons are bound to atomic nuclei by electromagnetic interactions
of widely varying strength.
Inner electrons are tightly bound to the nucleus and form ion cores.
Some outer electrons may be traded or shared between neighboring ion cores
in ionic or valence bonds, respectively.
In metallic bonds some outer electrons are mobilized.

Sorting out the diverse, complex, and interrelated phenomena is challenging.
The common strategy of many approaches is to transform specific aspects of the strongly
interacting ion cores and electrons into more weakly interacting collective modes.
The latter share many attributes with fundamental particles including
energy-momentum relations, spin, and exclusion statistics.
They scatter off each other elastically or inelastically, form bound states, or decay into
other modes.
The goal of transforming the strongly coupled constituent particles of condensed matter
into collective modes that behave like free particles is elusive except under idealized
circumstances related to dynamic or kinematic restrictions.

Harmonic lattice vibrations have linear equations of motion.
The collective modes exist in superpositions of infinite lifetimes without scattering.
Phonons have bosonic statistics.
Linear combinations of atomic orbitals produce fermionic counterparts: band electrons \cite{AM76}.
In a model that constrains the kinematics of collective modes
to one dimension and limits their dynamics to elastic two-body scattering, the momenta
and energies are conserved individually.
The collective modes are particles with exclusion statistics determined
by the (factorizing) $S$-matrix, analyzed via Bethe ansatz \cite{KBI93,Taka99,Suth04,EFG+05}.
Lattice degrees of freedom that are coupled by commuting operators are static.
They can be assembled into particles that are free and floating.
These particles have definite energies independent of the neighborhood.
Their mutual exclusion statistics are exotic \cite{LVP+08,LMK09,copic}.

This project imposes the kinematic constraint of one dimension and the dynamic
constraint of commuting operators.
It aims at shedding light on the assembly of structures from interactions and on the emergence
of order from these structures within that limited realm.
We expect that the results will open doors for situations
where the kinematic or dynamic constraints are relaxed.
The Ising chain for commuting spin operators ${S_l^z=\pm\frac{1}{2}}$
with nearest-neighbor (\textsf{nn}) coupling $J$, next-nearest-neighbor
(\textsf{nnn}) coupling $L$, magnetic field $h$, and periodic boundary conditions \cite{IT89},
\begin{equation}\label{eq:1}
\mathcal{H}=\sum_{l=1}^N\Big[JS_l^zS_{l+1}^z +LS_l^zS_{l+2}^z-hS_l^z\Big],
\end{equation}
is a suitable starting point from which it is possible to develop the methodology.
The spectrum consists of product states,
$|\sigma_1\cdots\sigma_N\rangle$ with $\sigma_l=\uparrow,\downarrow$.
The notation $|\sigma_1\cdots\sigma_N\rangle_p$ refers to sets of $p$ product states
(of periodicity $p$) that transform into each other via translations.

\begin{figure}[htb]
  \centering
  \includegraphics[width=9cm]{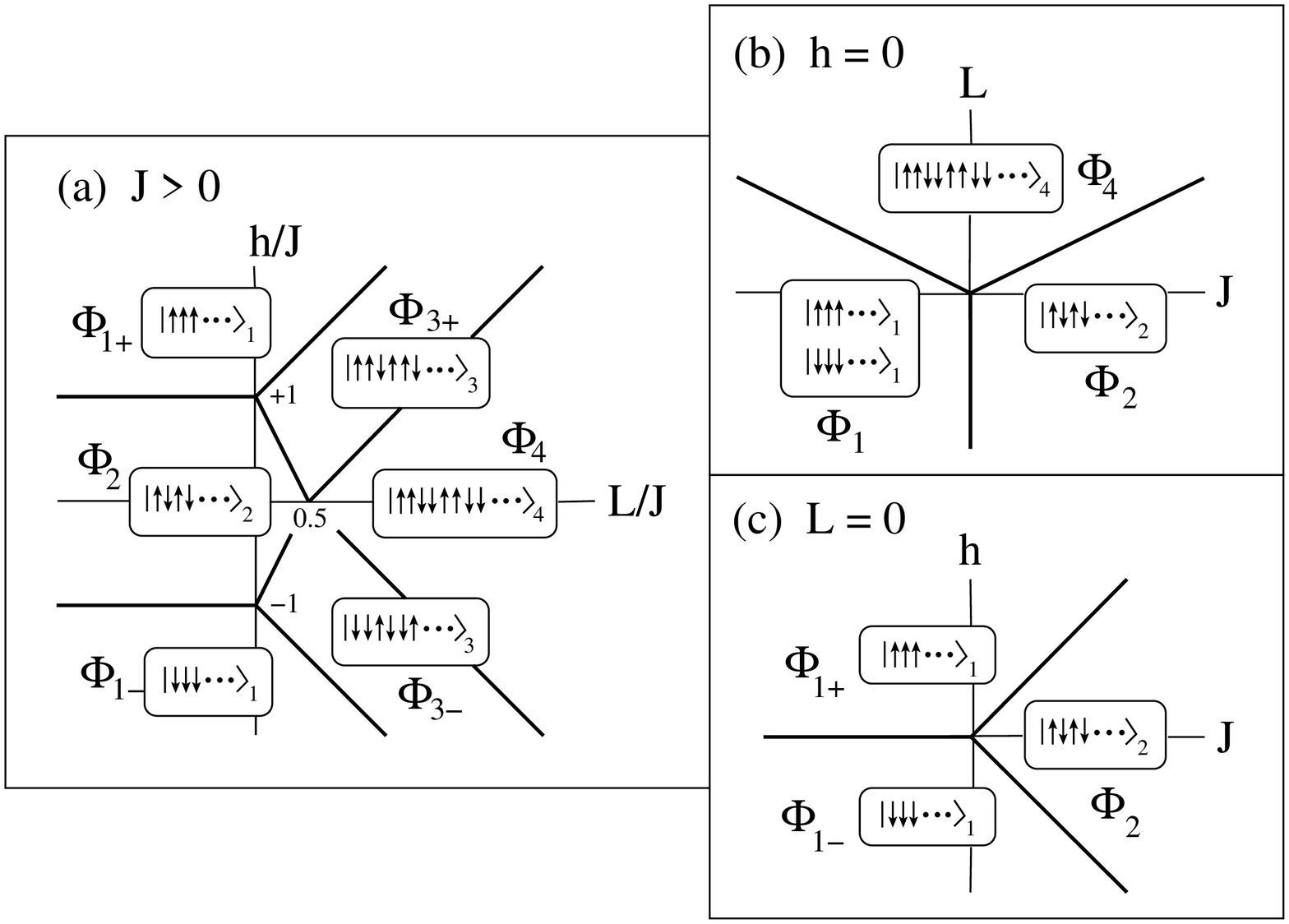}
  \caption{$T=0$ phase diagram (a) for positive nn coupling $J$ and
  nnn coupling $L$ of either sign in the parameter plane of scaled nnn-coupling $L/J$
  and scaled magnetic field $h/J$, (b) for $h=0$ in the $(J,L)$-plane, and
  (c) for $L=0$ in the $(J,h)$-plane. The text refers to phases and regions in parameter space by
  the same names.}
  \label{fig:plj5f1c}
\end{figure}

The zero-temperature phase diagram at $h=0$ features three phases with periodicities
$p=1,2,4$ (see figure~\ref{fig:plj5f1c}).
This includes phase $\Phi_1$ with \textsf{nn} spins aligned, phase $\Phi_2$ with \textsf{nn}
spins anti-aligned, and phase $\Phi_4$ (at $L>\frac{1}{2}|J|$) with \textsf{nn} spins alternatingly
aligned and anti-aligned.
The \textsf{nnn} spins are uniformly aligned in phases $\Phi_1,\Phi_2$ and uniformly
anti-aligned in phase $\Phi_4$.
At $h\neq0$ and $J>0$, phases $\Phi_2$ and $\Phi_4$ persist, phase $\Phi_1$ is split up
into two phases $\Phi_{1\pm}$, and two new plateau phases $\Phi_{3\pm}$ with periodicity
$p=3$ are stabilized.\footnote{The $h\neq0$ phase diagram is much simpler at $J<0$.
Only three of the phases exist: $\Phi_4$ in the sector $L/|J|>|h/J|+1/2$ and
$\Phi_{1\pm}$ at $h\gtrless0$ in the remaining sector of the $(L/|J|,h/|J|)$-plane.}
The $h\neq0$ phase diagram at $L=0$ features phases $\Phi_2$ in the sector $J>|h|$ and
$\Phi_{1\pm}$ at $h\gtrless0$ in the other sectors.

In section~\ref{sec:motspecat} we configure the physical vacua $\Phi_{1+}$, $\Phi_2$,
and $\Phi_4$ as the pseudo-vacua of sets of statistically interacting particles.
Salient features of the statistical mechanics of these particles are highlighted in section~\ref{sec:entlan}
(entropy, ordering tendencies) and section~\ref{sec:statmech} (populations in competition).
The emergence of structures from interactions on two levels is further discussed in
section~\ref{sec:strucint}.

%
\section{Motifs, species, categories}\label{sec:motspecat}
%
\emph{Motif} is a term borrowed from music, literature, and visual arts, where it refers to
fragments, themes, or patterns.
Now well established in the natural sciences, it is used in biochemistry, for example, to describe
patterns of nucleotides (codons) in DNA sequences.
Motifs as a representation of particles in many-body systems are common in statistical
mechanics \cite{Taka99,LMK09,HH93}.
In the present context, motifs are patterns of consecutive site variables $\sigma_l$ in Ising
product states.
Each motif characterizes either a particle of a particular species or an element of pseudo-vacuum.
Our goal is to find motifs representing particles that have a definite energy irrespective
of their location on the lattice relative to other particles.
This requirement limits the number of ways particles can be identified.
Every product eigenstate $|\sigma_1\cdots\sigma_N\rangle$ thus becomes a string of motifs.

Successive motifs interlink in a characteristic manner that depends on the range of interaction.
(i) The $h$-term in Hamiltonian (\ref{eq:1}) has zero range.
It permits the use of motifs that interlink by sharing no site variables
(e.g. $\uparrow\!+\!\uparrow=\uparrow\uparrow$).
All on-site energies are additive.
(ii) The $J$-term has range one.
It demands the use of motifs that interlink in one shared site variable
(e.g. $\uparrow\!\bar{\uparrow}+\bar{\uparrow}\!\uparrow=\uparrow\!\bar{\uparrow}\!\uparrow$).
This makes all \textsf{nn}-bond energies additive and the additivity of on-site energies can be maintained
by specific rules.
(iii) The $L$-term has range two, which requires that motifs interlink in two shared site variables
(e.g. $\uparrow\!\bar{\uparrow}\,\bar{\uparrow}\!+\!\bar{\uparrow}\,\bar{\uparrow}\!\uparrow
=\uparrow\!\bar{\uparrow}\,\bar{\uparrow}\!\uparrow$).
This guarantees that all \textsf{nnn}-bond energies are additive without jeopardizing the additivity of
\textsf{nn}-bond and on-site energies.
Longer-range couplings demand a more extensive overlap of motifs and more elaborate
rules for ensuring additivity of energy contributions from all Hamiltonian terms.

The search for motifs of a set of free particles that are excited from a given
pseudo-vacuum is guided by the additional optimization criteria that aim for the fewest
and shortest motifs.
In the context of a spin-1 Ising chain with \textsf{nn}-coupling we identified sets of six particles
excited from twofold pseudo-vacua and sets of seven particles from non-degenerate
pseudo-vacua \cite{copic}.
Here we use the same strategy to identify three sets of particles that generate the full spectrum
of (\ref{eq:1}) from pseudo-vacua $|\uparrow\uparrow\cdots\rangle_1$,
${|\uparrow\downarrow\uparrow\downarrow\cdots\rangle_2}$,
${|\uparrow\uparrow\downarrow\downarrow\uparrow\uparrow\downarrow\downarrow\cdots\rangle_4}$,
associated with the ground states at $h=0$.

The taxonomy of particles defined by their motifs involves \emph{structures} as sorted into
\emph{species} and \emph{functions} as emerging from \emph{categories}.
These features were the focus of reference~\cite{copic}.
In the present context the species will be very different but the categories will remain the same.
We shall again encounter \emph{compacts}, \emph{hosts}, \emph{tags}, \emph{hybrids},
and no further categories.
Compacts and hosts float in segments of pseudo-vacuum, tags are located inside hosts,
and hybrids are tags with hosting capability.
Particles from the same category but with different structures may collectively allow the emergence
of new functions.

The number of product eigenstates that contain specific numbers $\{N_m\}$ of particles from
all species of a given set is expressible by a multiplicity function $W(\{N_m\})$.
Its general structure, developed in the context of reference~\cite{copic}, remains operational
 without modification:
 \begin{subequations}\label{eq:2}
\begin{align}
W(\{N_m\})&=\frac{n_{\mathrm{pv}}N}{N-N^{(\alpha)}} \prod_{m=1}^M
\left( \begin{tabular}{c}
$d_m+N_m-1$ \\ $N_m$
\end{tabular} \right), \qquad N^{(\alpha)}=\sum_{m=1}^M\alpha_mN_m\,, \label{eq:2a} \\
d_m&=A_m-\sum_{m'=1}^{M}g_{mm'}(N_{m'}-\delta_{mm'}), \label{eq:2b}
\end{align}
\end{subequations}
where $n_{\mathrm{pv}}$ is the multiplicity of the pseudo-vacuum, the $A_m$ are capacity constants,
the $\alpha_m$ are size constants, and the $g_{mm'}$ are statistical interaction coefficients.
The generalized Pauli principle proposed by Haldane \cite{Hald91a} is encoded in (\ref{eq:2b})
with $d_m$ counting the number of open slots for particles of species $m$ in the presence of
$N_{m'}$ particles from any species $m'$, thus encapsulating the essence of statistical interaction.
All product states with particle content $\{N_m\}$ have energy
\begin{equation}\label{eq:3}
E\big(\{N_m\}\big)=E_{\mathrm{pv}}+\sum_{m=1}^M N_m\epsilon_m\,,
\end{equation}
where $\epsilon_m$ is the energy of particles from species $m$ relative to the pseudo-vacuum,
which has (absolute) energy $E_{\mathrm{pv}}\,$.

\subsection{Particles generated from
$|\uparrow\downarrow\uparrow\downarrow\cdots\rangle_2$}\label{sec:nnnp1}
\begin{figure}[b]
  \centering
  \includegraphics[width=9cm]{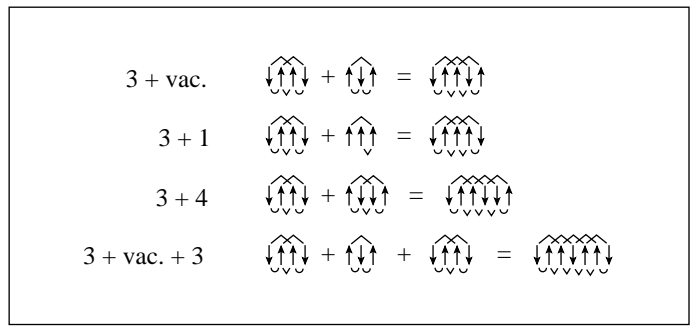}
  \caption{Interlinking elements of pseudo-vacuum and particles from
    table~\ref{tab:specsP1}. Some \textsf{nn} bonds $(\cup)$ are counted
    half, others $(\vee)$ full. All \textsf{nnn} bonds $(\wedge)$
    are counted fully.}
  \label{fig:plj5f3}
\end{figure}

The physical vacuum in region $\Phi_2$ is the twofold N\'eel state
$|\uparrow\downarrow\uparrow\cdots\rangle_2$, here selected as the pseudo-vacuum for
$M=4$ species of particles with specifications as compiled in table~\ref{tab:specsP1} .
The permissible configurations of particles from these species generate the complete spectrum
of $\mathcal{H}$.
The motifs interlink as illustrated in figure~\ref{fig:plj5f3}.
Particles $m=3,4$ are hosts and particles $m=1,2$ tags.
Hosts can be placed into segments of pseudo-vacuum and tags inside hosts.
In this instance, hosts 3 accommodate tags 1 only and hosts 4 tags 2 only.

\begin{table}[htb]
\vspace{-3mm}
  \caption{Specifications of $M=4$ species of particles excited from the N{\'e}el state $(n_{\mathrm{pv}}=2)$
  ${|\uparrow\downarrow\uparrow\cdots\rangle_2}$: motif, category, species,
    energy (relative to pseudo-vacuum), spin, capacity constants, size constants (left), and
    statistical interaction coefficients (right). Segments of $\ell$ vacuum elements,
    $\uparrow\downarrow\uparrow,
    \downarrow\uparrow\downarrow$, have energy $\ell(L-J)/4$. At $h\neq0$ the entries of
    $\epsilon_m$ must be amended by $-s_mh$.}\label{tab:specsP1}  \vspace{2ex}
\begin{center}
\begin{tabular}{|ccc||cccc|}
\hline
motif & category &$m$ & $\epsilon_{m}$ & $s_m$ & $A_{m}$ & $\alpha_m$
\\ \hline \hline\rule[-2mm]{0mm}{6mm}
$\uparrow\uparrow\uparrow$ & tag & $1$ & $\frac{1}{2}J$
& $+\frac{1}{2}$ &$0$ & $1$ \\ \rule[-2mm]{0mm}{5mm}
$\downarrow\downarrow\downarrow$ & tag & $2$ & $\frac{1}{2}J$
& $-\frac{1}{2}$ & $0$ & $1$ \\ \rule[-2mm]{0mm}{5mm}
$\downarrow\uparrow\uparrow\downarrow$ & host & $3$ &
$\frac{1}{2}J-L$ & $+\frac{1}{2}$ & $\frac{N-1}{2}$ & $1$ \\ \rule[-2mm]{0mm}{5mm}
$\uparrow\downarrow\downarrow\uparrow$ & host & $4$ &
$\frac{1}{2}J-L$ & $-\frac{1}{2}$ & $\frac{N-1}{2}$ & $1$\\
\hline
\end{tabular}\hspace{7mm}%
%
\begin{tabular}{|c||rrrr|}
\hline
$g_{mm'}$ & $1$ & $2$ & $3$ & $4$
\\ \hline \hline\rule[-2mm]{0mm}{6mm}
$1$ & $~~0$ & $~~0$ & $-1$ & $0$ \\ \rule[-2mm]{0mm}{5mm}
$2$ & $0$ & $0$ & $0$ & $-1$  \\ \rule[-2mm]{0mm}{5mm}
$3$ & $\frac{1}{2}$ & $\frac{1}{2}$ & $\frac{3}{2}$ & $\frac{1}{2}$
\\ \rule[-2mm]{0mm}{5mm}
$4$ & $\frac{1}{2}$ & $\frac{1}{2}$ & $\frac{1}{2}$ & $\frac{3}{2}$\\
\hline
\end{tabular}
\end{center}
\end{table}

The energies of hosts and tags are calculated differently.
In hosts (or elements of vacuum) we count interior \textsf{nn} bonds fully but the outermost
\textsf{nn} bonds only half.
In tags we do not count the \textsf{nn} bond on the left but do count the one on the right fully.
All \textsf{nnn} bonds are counted fully in each particle (or element of vacuum).
These rules are implemented in the entries for $\epsilon_m$.
In analogous manner a spin $s_m$ can be assigned to particles from each species.

Tags have vanishing $A_m$ \cite{copic}.
Open slots for tags are created by the placement of hosts.
The number of tags that can latch on to a given host is only limited by the space to which the latter
can expand in the process.
In one product state tags (of one species) are lined up around the chain with no need of a host.

The $g_{mm'}$ from table~\ref{tab:specsP1} as used in (\ref{eq:2b}) may be interpreted as follows:
(i) Adding a tag does not change the capacity for further tags.
Any slot taken by a tag opens up exactly one slot for a tag of the same kind.
(ii) In the process of adding a tag its host must expand, which reduces the
capacity for further hosts of either kind in the same manner.
(iii) Adding a host increases the capacity for tags that it can host and
has no effect on the capacity for tags it cannot host.
(iv) Adding a host of any kind diminishes the capacity for further hosts of the same
kind more strongly than for hosts of the other kind.
Different hosts can be interlinked directly, whereas identical hosts must be spaced by at least one
element of vacuum.

The two vectors of phase $\Phi_2$ are the pseudo-vacuum of the particles considered here.
The two vectors of phase $\Phi_1$ are each a solid of $N$ close-packed tags of one kind.
By contrast, the four vectors of phase $\Phi_4$ are each a solid of $\frac{1}{2}N$ hosts
in an alternating pattern.

For the case of vanishing \textsf{nnn}-coupling $(L=0)$ the four species of particles from
table~\ref{tab:specsP1}  can be replaced by the two species in table~\ref{tab:specsol}.
The motifs are shorter and interlink in one shared site variable.
All \textsf{nn} bonds are now counted fully and all sites half in each motif.
The new particles (named solitons) are well known from previous work \cite{LVP+08,VIS}.
Their statistical interaction is semionic, an attribute shared with the spinons identified in the
$XXZ$ and Haldane-Shastry models \cite{Hald91,KMW08}.
In the taxonomy of reference~\cite{copic} they are compacts.
In figure~\ref{fig:plj5f1c}~(c) phase $\Phi_2$ is the pseudo-vacuum of solitons whereas phase $\Phi_{1+}$
$(\Phi_{1-})$ is a solid of $N$ spin-up (spin-down) solitons.

\begin{table}[htb]
  \caption{Specifications of $M=2$ species of particles excited from the N{\'e}el state  for $L=0$
  ($n_{\mathrm{pv}}=2$) \mbox{$|\uparrow\downarrow\uparrow\downarrow\cdots\rangle_2$}.
  Segments of $\ell$ vacuum elements, $\uparrow\downarrow,
  \downarrow\uparrow$, have energy $-\ell J/4$. At $h\neq0$ the entries of $\epsilon_m$ must be
  amended by $-s_mh$.}\label{tab:specsol}  \vspace*{2mm}
\begin{center}
\begin{tabular}{|ccc||cccc|}
\hline motif & cat. & $m$ & $\epsilon_{m}$ & $s_{m}$ & $A_{m}$ & $\alpha_m$
\\ \hline\hline  \rule[-2mm]{0mm}{6mm}
$\uparrow\uparrow$ & comp. & $+$ & $\frac{J}{2}$ & $+\frac{1}{2}$ & $\frac{N-1}{2}$ & $1$
\\ \rule[-2mm]{0mm}{5mm}
$\downarrow\downarrow$ & comp. & $-$ & $\frac{J}{2}$ & $-\frac{1}{2}$ & $\frac{N-1}{2}$ & $1$
\\\hline
\end{tabular}\hspace{8mm}
\begin{tabular}{|c||rr|}
\hline
$g_{mm'}$ & $+$ & $-$  \\\hline  \hline \rule[-2mm]{0mm}{6mm}
$+$ & $~~\frac{1}{2}$ & $~~\frac{1}{2}$ \\ \rule[-2mm]{0mm}{5mm}
$-$ & $\frac{1}{2}$ & $\frac{1}{2}$\\
\hline
\end{tabular}
\end{center}
\end{table}

The motifs of solitons are not fragments of the original motifs.
They interlink differently.
At $L\neq0$ we have two spin-up particles (tag 1 and host 3)
and two spin-down particles (tag 2 and host 4).
At $L=0$ the energies of both spin-up (or spin-down) particles become equal.
In this case we can get away more economically with one spin-up and one
spin-down soliton.
Returning to $L\neq0$, the energy of a soliton depends on its position relative to other solitons.
The interaction energy is eliminated by switching back to the extended set of four species.

\subsection{Particles generated from
$|\uparrow\uparrow\cdots\rangle_1$}\label{sec:nnnp0}
Phase $\Phi_1$ comprises the twofold spin-polarized state
$|\uparrow\uparrow\uparrow\cdots\rangle_1$, $|\downarrow\downarrow\downarrow\cdots\rangle_1$.
Here we adopt the first vector as the pseudo-vacuum for $M=5$ species of particles with
specifications compiled in table~\ref{tab:specsP0}.
\begin{table}[!b]
  \caption{Specifications of $M=5$ species of particles excited from the spin-polarized state
  $(n_{\mathrm{pv}}=1)$   \mbox{$|\uparrow\uparrow\cdots\rangle_1$}.
  Segments of $\ell$ vacuum elements, $\uparrow\uparrow\uparrow$,
  have energy $\ell(L+J)/4$. At $h\neq0$ the entries of $\epsilon_m$ must be amended by
  $-s_mh$.}\label{tab:specsP0}   \vspace*{2mm}
\begin{center}
\begin{tabular}{|ccc||cccc|}
\hline
motif & category & $m$ & $\epsilon_{m}$ & $s_m$ & $A_{m}$ & $\alpha_m$
\\ \hline\hline  \rule[-2mm]{0mm}{6mm}
$\uparrow\uparrow\downarrow\downarrow\uparrow\uparrow$ & host
& $1$ & $-J-2L$ & $-2$ & $N-3$ & $3$ \\ \rule[-2mm]{0mm}{5mm}
$\uparrow\uparrow\downarrow\uparrow\uparrow$ & host
& $2$ & $-J-L$ & $-1$ & $N-2$ & $2$ \\ \rule[-2mm]{0mm}{5mm}
$\downarrow\downarrow\downarrow$ & tag
& $3$ & $0$ & $-1$ & $0$ & $1$ \\ \rule[-2mm]{0mm}{5mm}
$\uparrow\downarrow\uparrow\downarrow,\downarrow\uparrow\downarrow\uparrow$ & tag & $4$ &
$-J$ & $-1$ & $0$ & $2$ \\ \rule[-2mm]{0mm}{5mm}
$\downarrow\downarrow\uparrow\downarrow\downarrow$ & hybrid &$5$ &
$-J-L$ & $-2$ & $0$ & $3$
\\ \hline
\end{tabular}

\vspace*{4ex}
\begin{tabular}{|c||rrrrr|}
\hline
$g_{mm'}$ & $1$ & $2$ & $~~3$ & $~~4$ & $5$
\\ \hline \hline \rule[-2mm]{0mm}{6mm}
$1$ & $4$ & $3$ & $1$ & $2$ & $3$\\ \rule[-2mm]{0mm}{5mm}
$2$ & $3$ & $3$ & $1$ & $2$ & $3$\\ \rule[-2mm]{0mm}{5mm}
$3$ & $-1$ & $0$ & $0$ & $0$ & $-1$\\ \rule[-2mm]{0mm}{5mm}
$4$ & $-2$ & $-1$ & $0$ & $0$ & $-1$\\ \rule[-2mm]{0mm}{5mm}
$5$ & $-1$ & $0$ & $0$ & $0$ & $0$
\\
\hline
\end{tabular}
\end{center}
\end{table}
Three categories are represented.
Host 2 accommodates only tag 4 whereas host 1 accommodates both tags and the hybrid.
The hybrid, in turn, is capable of hosting both tags.
The energies of hosts and tags are calculated as in section~\ref{sec:nnnp1}.
The rules for hybrids are the same as those for tags.
The pseudo-vacuum is spin-polarized and $s_m$ is not a spin in the usual sense.
It enables us to write the magnetic-field contribution to $\epsilon_m$ in the form $-s_mh$.

\begin{figure}[htb]
  \centering
  \includegraphics[width=125mm]{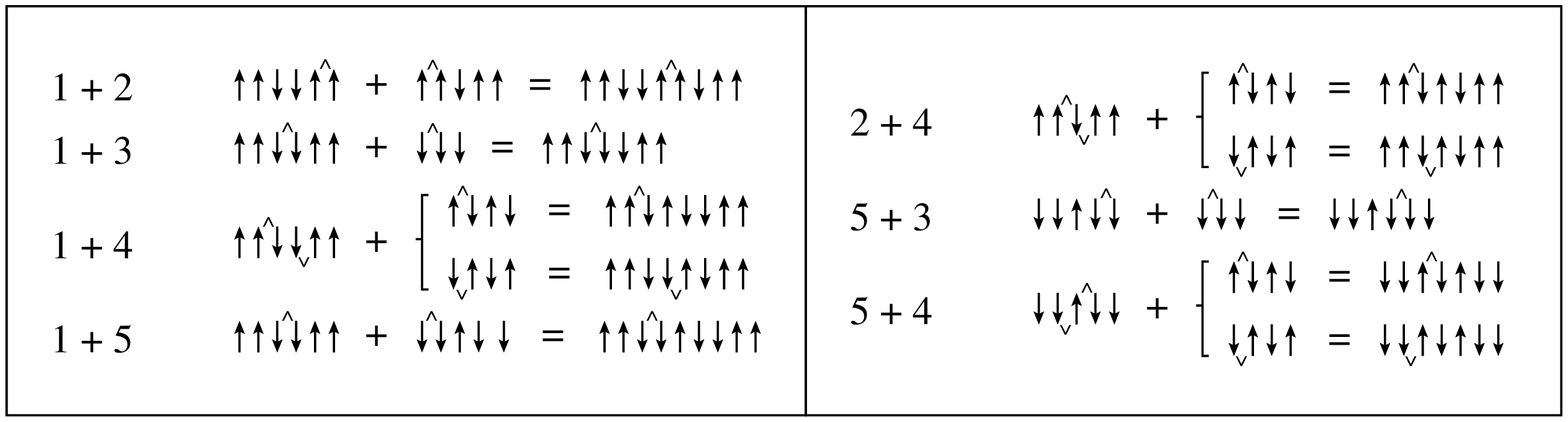}
  \caption{Interlinking particles from table~\ref{tab:specsP0}.
  The number of \textsf{nnn} bonds is preserved and the number of \textsf{nn} bonds
    is reduced by one. The shared bond is marked by $\wedge$ or $\vee$.}
  \label{fig:plj5f3m}
\end{figure}

Various combinations of hosts, tags, and hybrids are illustrated in figure~\ref{fig:plj5f3m}.
In the search for particles that are free of interaction energies, attention had to be paid
to the requirement that the implantation of a tag or hybrid into a host (or a tag into hybrid) leaves
the sums of aligned and anti-aligned \textsf{nnn} bonds invariant.

The hosting capabilities of particles $m'=1,2,5$ are encoded in negative statistical interaction
coefficients $g_{mm'}$.
In all instances except one, we have $g_{mm'}=-1$.
The fact that host 1 has two interior slots to accommodate tag 4 requires that $g_{41}=-2$.
Tag 3 has zero energy, a consequence of our choice of pseudo-vacuum.
The physical vacuum in region $\Phi_1$ comprises the pseudo-vacuum
$|\uparrow\uparrow\cdots\rangle_1$ and a solid of tags 3,
$|\downarrow\downarrow\cdots\rangle_1$.
In region $\Phi_2$ the physical vacuum is a solid of negative-energy tags 4 (two vectors)
and in region $\Phi_4$ a solid of negative-energy hosts 1 (four vectors).

For the case $L=0$ we can again get away with fewer particles that have shorter motifs and
interlink with shorter overlap: one host and one tag as listed in table~\ref{tab:specht}.
In figure~\ref{fig:plj5f1c}~(c) phase $\Phi_{1+}$ is the pseudo-vacuum,
phase $\Phi_{1-}$ is a solid of tags, and phase $\Phi_2$ is a solid of hosts.
The two species of table~\ref{tab:specht} are free of \textsf{nn} interaction energies.
The \textsf{nnn} interaction energies between them can be eliminated if we allow them to
assemble into the five species of table~\ref{tab:specsP0}.

\begin{table}[h]
  \caption{Specifications of $M=2$ species of particles excited from the spin-polarized state for
  $L=0$   ($n_{\mathrm{pv}}=1$) $|\uparrow\uparrow\cdots\rangle_1$.
  Segments of $\ell$ vacuum elements, $\uparrow\uparrow$, have energy $\ell J/4$.
  At $h\neq0$ the entries of $\epsilon_m$ must be amended by
  $-s_mh$.}\label{tab:specht} \vspace*{2mm}
\begin{center}
\begin{tabular}{|ccc||cccc|}
\hline
motif & cat. & $m$ & $\epsilon_{m}$ & $s_{m}$ & $A_{m}$ & $\alpha_m$
\\ \hline\hline  \rule[-2mm]{0mm}{6mm}
$\uparrow\downarrow\uparrow$ & host & $H$ & $-J$ & $-1$ & $N-1$ & $1$
\\ \rule[-2mm]{0mm}{5mm}
$\downarrow\downarrow$ & tag & $T$ & $0$ & $-1$ & $0$ & $1$
\\\hline
\end{tabular}\hspace{8mm}
\begin{tabular}{|c||rr|}
\hline
$g_{mm'}$ & $H$ & $T$  \\ \hline \hline \rule[-2mm]{0mm}{6mm}
$H$ & $2$ & $~~1$ \\ \rule[-2mm]{0mm}{5mm}
$T$ & $-1$ & $0$
\\\hline
\end{tabular}
\end{center}
\end{table}

\subsection{Particles generated from
$|\uparrow\uparrow\downarrow\downarrow\uparrow\uparrow\cdots\rangle_4$}\label{sec:nnnp2}
In region $\Phi_4$ the fourfold state
$|\uparrow\uparrow\downarrow\downarrow\uparrow\uparrow\cdots\rangle_{4}$
is the physical vacuum.
Our search for free particles that generate the entire spectrum from this state
configured as pseudo-vacuum has produced $M=4$ compacts with motifs and specifications
compiled in table~\ref{tab:specsP2}.
All particles or elements of pseudo-vacuum comprise two \textsf{nn} bonds and one
\textsf{nnn} bond.
They again interlink by sharing one \textsf{nn} bond.
Their energy content consists of one half of each \textsf{nn} bond energy
plus the full \textsf{nnn} bond energy.

\begin{table}[htb]
  \caption{Specifications of $M=4$ species of particles excited from the state $(n_{\mathrm{pv}}=4)$
  ${|\uparrow\uparrow\downarrow\downarrow\uparrow\uparrow\cdots\rangle_4}$.
  Segments of $\ell$ vacuum elements $\uparrow\uparrow\downarrow,
  \uparrow\downarrow\downarrow, \downarrow\downarrow\uparrow, \downarrow\uparrow\uparrow$
  have energy $-\ell L/4$. At $h\neq0$ the entries of $\epsilon_m$ must be amended by
  $-s_mh$.}\label{tab:specsP2}  \vspace*{2mm}
\begin{center}
\begin{tabular}{|ccc||cccc|}
\hline
motif & category & $m$ & $\epsilon_{m}$ & $s_m$ & $A_{m}$ & $\alpha_m$
\\ \hline\hline  \rule[-2mm]{0mm}{6mm}
$\uparrow\uparrow\uparrow$ & compact & $1$ & $\frac{1}{4}(2L+J)$
& $+\frac{1}{2}$ & $\frac{N-1}{4}$ & $1$ \\ \rule[-2mm]{0mm}{5mm}
$\downarrow\downarrow\downarrow$ & compact & $2$
& $\frac{1}{4}(2L+J)$ & $-\frac{1}{2}$ & $\frac{N-1}{4}$ & $1$ \\ \rule[-2mm]{0mm}{5mm}
$\uparrow\downarrow\uparrow$ & compact & $3$ & $\frac{1}{4}(2L-J)$ & $+\frac{1}{2}$ &
$\frac{N+1}{4}$ & $1$ \\ \rule[-2mm]{0mm}{5mm}
$\downarrow\uparrow\downarrow$ & compact & $4$ & $\frac{1}{4}(2L-J)$ & $-\frac{1}{2}$ &
$\frac{N+1}{4}$ & $1$
\\\hline
\end{tabular}\hspace{3mm}%
%
\begin{tabular}{|c||rrrr|}
\hline
$g_{mm'}$ & $~~1$ & $~~2$ & $3$ & $4$
\\\hline  \hline \rule[-2mm]{0mm}{6mm}
$1$ & $\frac{1}{4}$ & $\frac{1}{4}$ & $-\frac{1}{4}$ & $\frac{3}{4}$ \\
\rule[-2mm]{0mm}{5mm}
$2$ & $\frac{1}{4}$ & $\frac{1}{4}$ & $\frac{3}{4}$ & $-\frac{1}{4}$  \\
\rule[-2mm]{0mm}{5mm}
$3$ & $\frac{1}{4}$ & $\frac{1}{4}$ & $\frac{3}{4}$ & $-\frac{1}{4}$  \\
\rule[-2mm]{0mm}{5mm}
$4$ & $\frac{1}{4}$ & $\frac{1}{4}$ & $-\frac{1}{4}$ & $\frac{3}{4}$
\\\hline
\end{tabular}
\end{center}
\end{table}

The assignment of a spin $\pm\frac{1}{2}$ to the four particles is based on the following reasoning.
Take two interlinked elements of pseudo-vacuum with zero spin,
$\uparrow\uparrow\downarrow\downarrow$ or
$\downarrow\downarrow\uparrow\uparrow$, add one of the four particles, and check the spin
of the resulting entity. For example,
$\uparrow\uparrow\downarrow\downarrow +\uparrow\uparrow\uparrow \hat{=}
\uparrow\uparrow\uparrow\downarrow\downarrow$ and
$\uparrow\uparrow\downarrow\downarrow +\uparrow\downarrow\uparrow \hat{=}
\uparrow\uparrow\downarrow\uparrow\downarrow$ produce entities with spin $+\frac{1}{2}$.

Four coefficients $g_{mm'}$ are negative.
Here we do not have the standard host-tag scenario described previously.
To understand the variant scenarios we note that each motif (four elements of vacuum and
four particle species) can be followed only by two out of eight motifs.
Only the motifs of particles 1,2 can follow themselves.
Thus adding a particle $m'=1$ or $m'=2$ merely reduces the total number of
open slots for further particles of any species.
This explains that all coefficients in the first two columns are equal and positive.
The coefficients in columns $m'=3,4$ reflect stronger exclusion in some instances
and accommodation in others, involving two distinct mechanisms.

\begin{itemize}
\item[(i)] We note that particles 3,4 accommodate each other mutually.
The addition of a particle $m'=3$ opens up a slot for a particle $m=4$ and vice versa.
This mutual accommodation is reflected in the coefficients $g_{34}=g_{43}=-1/4$.
At the same time the addition of a particle $m'=3$ or $m'=4$ closes down a slot for
further particles of the same species.
This accounts for the more strongly positive coefficients $g_{33}=g_{44}=3/4$.
\item[(ii)] The sequence between any motif and the motif of particle $m=1$ can always
be shortened by the insertion of a particle $m'=3$ wherever it fits.
By contrast, the insertion of a particle $m'=4$ wherever it fits will lengthen that sequence.
The presence of a particle $m'=3$ thus increases the capacity of the system for particles $m=1$
and the presence of a particle $m'=4$ has the opposite effect.
This is reflected in the coefficients $g_{13}=-1/4$, $g_{14}=3/4$.
Analogous reasoning explains the tabulated values of $g_{24}$ and $g_{23}$.
\end{itemize}

The ferromagnetic (FM) phase $\Phi_1$ consists of two states with broken spin-flip symmetry,
one being a solid of particles 1, the other a solid of particles 2.
The antiferromagnetic (AFM) phase $\Phi_2$ also consists of two states, each solid
composed of particles 3,4 in an alternating sequence with broken translational symmetry.
FM particles 1,2 and AFM particles 3,4 both have spin $s_m=\pm\frac{1}{2}$.
By interlinking differently, the former produce a uniform magnetization in phase $\Phi_1$ and
the latter a staggered magnetization in phase $\Phi_2$.

%
\section{Entropy landscapes}\label{sec:entlan}
%
The statistically interacting particles from the three sets identified in section~\ref{sec:motspecat}
have definite shapes and energies.
These floating objects are assembled from localized spins by the \textsf{nn}
and \textsf{nnn} couplings of Hamiltonian (\ref{eq:1}).
The Ising chain is an open system of particles with energies $\epsilon_m$ depending on $J,L,h$.
Here we abandon the Ising context and focus entirely on particles with interlinking motifs.

The statistical interaction between the particles depends on their shapes and on the nature
of the pseudo-vacuum, but not on the particle energies $\epsilon_m$.
It is instructive to explore the effects of statistical interactions produced by particular
shapes in a setting where particle energies do not factor in.
To this end we consider the configurational entropy as derived from the multiplicity expression
(\ref{eq:2}) for $N,N_m\gg1$ via $S=k_{\mathrm{B}}\ln W$:
\begin{subequations}\label{eq:snmp}
\begin{equation}
S(\{N_m\}) = k_{\mathrm{B}}\sum_{m=1}^M\Big[\big(N_{m}+Y_m\big)\ln\big(N_m+Y_m\big)
-N_m\ln N_m -Y_m\ln Y_m\Big],
\end{equation}
\begin{equation}
  Y_m \doteq A_m-\sum_{m'=1}^Mg_{mm'} N_{m'}\,.
\end{equation}
\end{subequations}
The functional dependence of $S$ on the populations $N_m$ of particle species from a given
set determines an entropy landscape shaped by
the statistical interactions alone.

If only one species is present, the statistical interaction reduces to an exclusion principle.
The function $S(N_m)$ vanishes identically for any species of tags or hybrids,
which have $A_m=0$ and $g_{mm}=0$.
These particles can only exist inside hosts.
Compacts and hosts exist in segments of pseudo-vacuum.
They have $A_m\propto N$ and $g_{mm}>0$.
The entropy $S(N_m)$ of a single species of hosts or compacts is nonzero for
$0<N_m<A_m/g_{mm}$ and zero at the endpoints.

The signature of the statistical interaction between any two species $m, m'$
is best visualized in a reduced entropy landscape, in contour plots
of the entropy per site, $\bar{S}(\bar{N}_m,\bar{N}_{m'})=S(N_m/N,N_{m'}/N)/N$
with the populations of all other species suppressed.
All thermodynamic processes described in the following are understood to be quasi-static
and to be implemented in an open system.\footnote{The Ising model (\ref{eq:1}) is one particular
realization with specific particle energies $\epsilon_m\,$.}
Equilibration would be problematic in closed, one-dimensional systems of particles from more
than one species.

\subsection{Hosts, tags, hybrids}\label{sec:hotahy}
Beginning with the four species from section~\ref{sec:nnnp1} we consider
the entropy landscapes pertaining to the two hosts, a host and the tag it does or does not
accommodate, and the two tags.
In each instance we vary the population densities
of two species over the permissible range while keeping the population densities of the
other two species constant at very low values.

\begin{figure}[htb]
  \includegraphics[width=70mm]{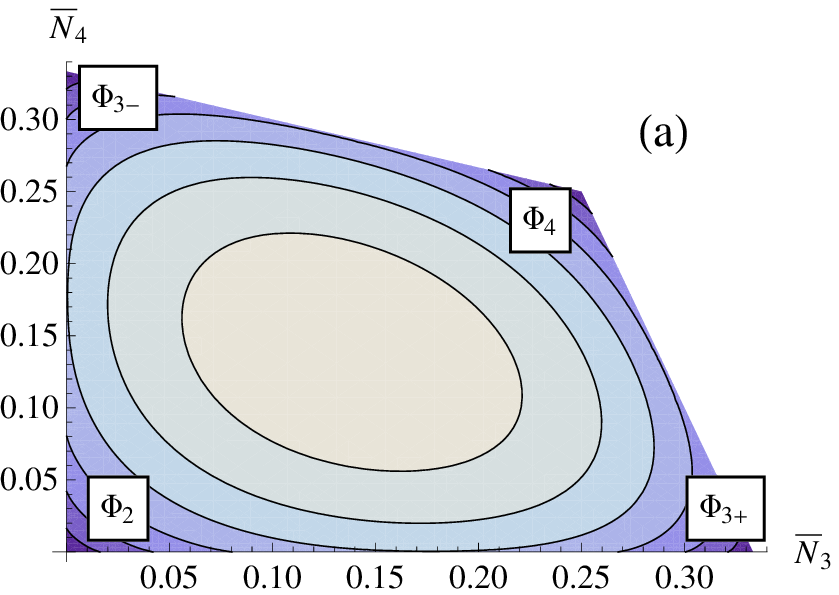}\hspace{5mm}%
  \includegraphics[width=70mm]{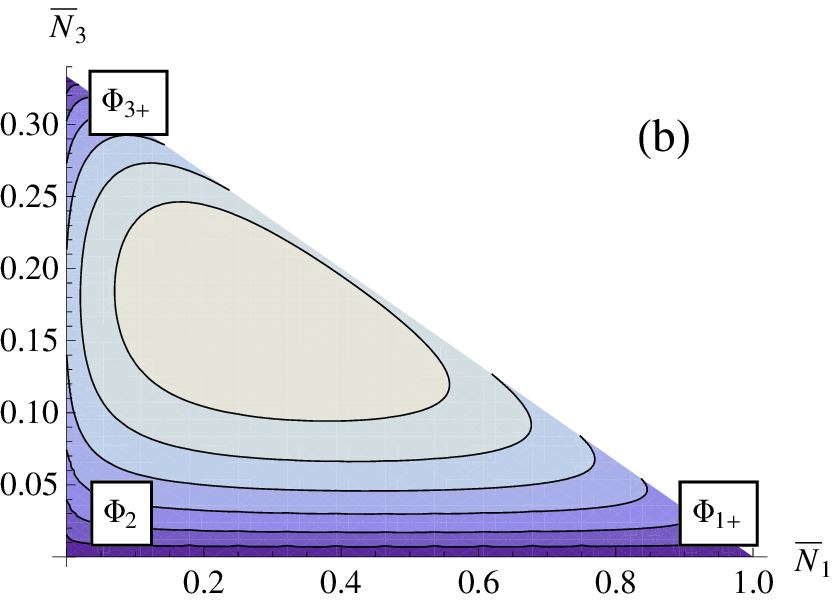}
  \includegraphics[width=70mm]{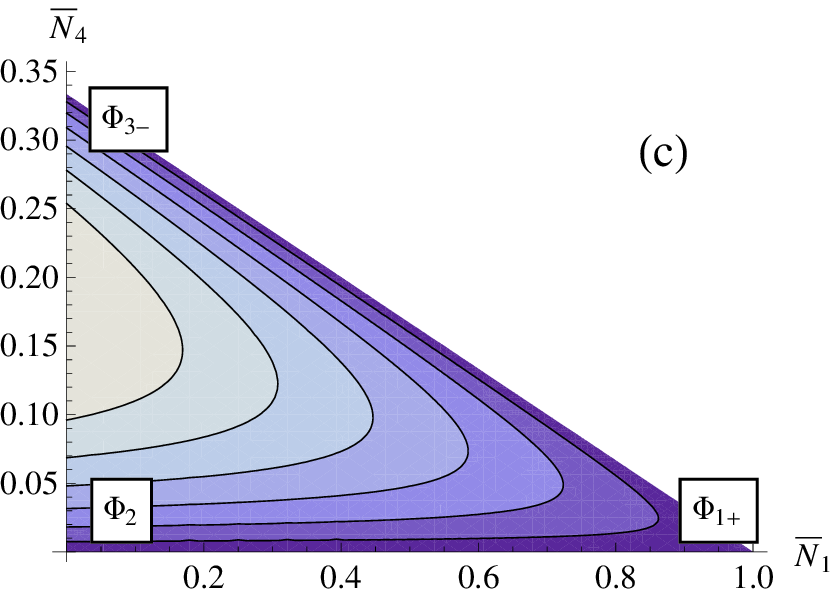}\hspace{5mm}%
  \includegraphics[width=70mm]{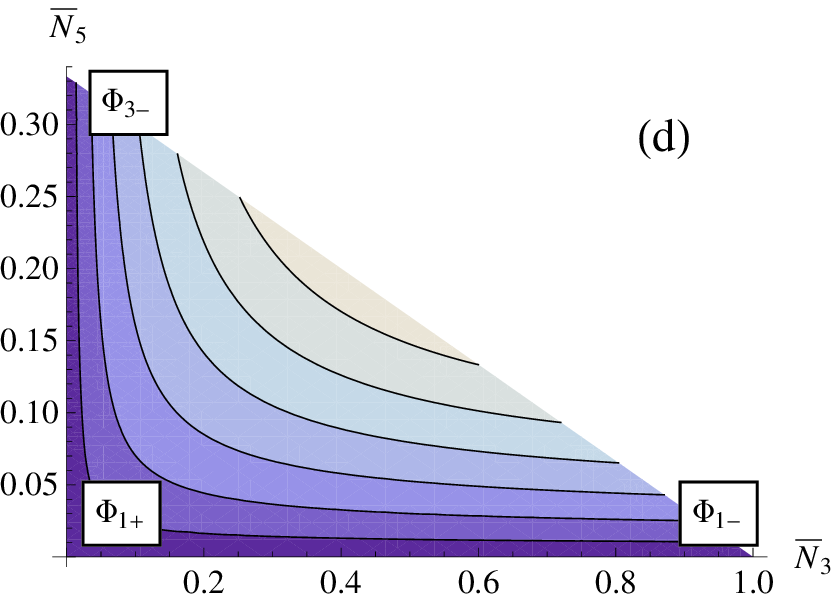}
\caption{(Color online) Entropy per site, $\bar{S}/k_{\mathrm{B}}$, versus population densities, $\bar{N}_m, \bar{N}_{m'}$,
of two species from (a)--(c) table~\ref{tab:specsP1} and (d) from table~\ref{tab:specsP0}.
The population densities of the other species are kept at $10^{-5}$.
The contours are at (a) $0.071\ell$, (b) $0.068\ell$, (c) $0.039\ell$, (d) $0.058\ell$, $\ell=1,\ldots,6$.}
  \label{fig:picnnn-1-enta}
\end{figure}

The entropy landscape of the two hosts as shown in figure~\ref{fig:picnnn-1-enta}~(a) has borders
of quadrilateral shape.
The four corners correspond to pure phases with vanishing entropy.
Hosts interlink with elements of pseudo-vacuum in multiple configurations.
This explains the nonzero entropy along the two sides on the axes.
The other two sides represent states that are crowded with hosts of both species.
If $\bar{N}_3=\bar{N}_4$ the state of maximum population density is unique.
Hosts 3,4 interlink directly.
Hosts 3,3 or 4,4 are spaced by at least one element of pseudo-vacuum.
The corner states $\Phi_{3\pm}$ contain $\frac{1}{3}N$ close-packed hosts from a single species.
The state $\Phi_4$ contains $\frac{1}{2}N$ close-packed hosts from both species arrayed in
an alternating sequence.
Heading from point $\Phi_{3+}$ toward point $\Phi_4$ involves the repeated replacement of one
host 3 by three hosts 4.

A qualitatively different entropy landscape pertains to host 3 and tag 1 (hosted by 3)
as shown in figure~\ref{fig:picnnn-1-enta}~(b).
The borders are of triangular shape.
Hosts 3 alone (vertical leg) generate entropy but tags 1 alone (horizontal leg) do not.
All tags 1 are arrayed uniformly inside hosts 3.
Only the expanded hosts of multiple sizes have positional disorder, not the tags inside.
As the hosts disappear, so does the entropy.
The tags contribute to the entropy only indirectly by expanding hosts to different sizes.
Close-packed configurations of hosts stuffed with tags are represented by points on the hypotenuse.
Near the middle the entropy is largest, generated by close-packed hosts expanded to many
different sizes.
Near one end, we have many more hosts but with only few small sizes represented.
Near the other end, we have few hosts expanded to large sizes.
Either trend reduces the entropy.

The entropy landscape is yet different for host 4 and tag 1 (not hosted by 4) as shown in
figure~\ref{fig:picnnn-1-enta}~(c).
All tags 1 are confined to very few hosts 3, incapable of producing any significant entropy by
themselves (horizontal leg).
Almost all hosts 4 have the minimal size due to the near absence of tags 2.
These hosts 4 produce positional entropy along the vertical leg.
The hypotenuse describes configurations of near uniform segments of hosts 4
separated by rare hosts 3 filled to various sizes with uniform arrays of tags 1.
This arrangement has very low entropy.
Starting with any density of hosts 4, the entropy always decreases when we add tags 1.
All tags 1 added replace elements of pseudo-vacuum and thus reduce the options for
positioning hosts 4.
The tags themselves are lumped together inside few hosts 3.

The two tags 1 and 2  have an entropy landscape (not shown)  that is also triangular
but very flat near zero entropy.
The two species of tags do not mix.
They are arrayed uniformly in separate hosts, present only in very low densities.

The entropy landscapes of the five species of hosts, tags, and hybrids from section~\ref{sec:nnnp0}
are all of triangular shape.
The hybrid 5 gives rise to one new feature, e.g. in combination with tag 3 as illustrated in
figure~\ref{fig:picnnn-1-enta}~(d).
Both species exist inside hosts 1 or 2.
Tags 3 alone or hybrids 5 alone produce no significant entropy for reasons already stated.
However, the two species can coexist inside the same host in many different configurations.
Almost all entropy is now generated inside the few hosts available.

In surveying the four panels of figure~\ref{fig:picnnn-1-enta} we see that the entropy
maximum is realized under diverse circumstances, with both species or only one species
present at intermediate density, or with two species in close-packed configurations.
As expected, maximum positional ordering of the particles (zero entropy) occurs
if no particles are present or if the system is close-packed with one species.
Close-packing with two species is also found to produce maximum positional order
either through alternate stacking [panel (a)] or through segregation [panel (c)].

\subsection{Compacts}\label{sec:cmpcts}
\begin{figure}[htb]
  \includegraphics[width=70mm]{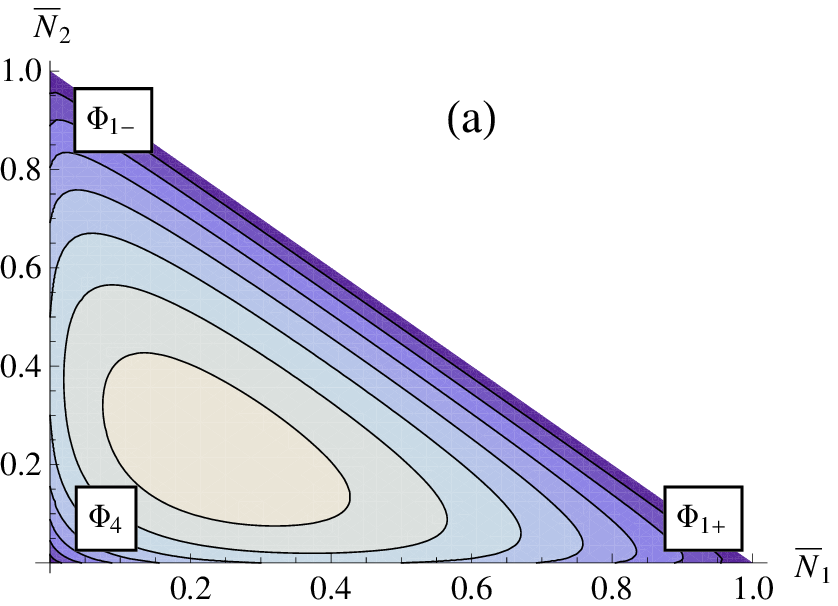}\hspace{5mm}%
  \includegraphics[width=70mm]{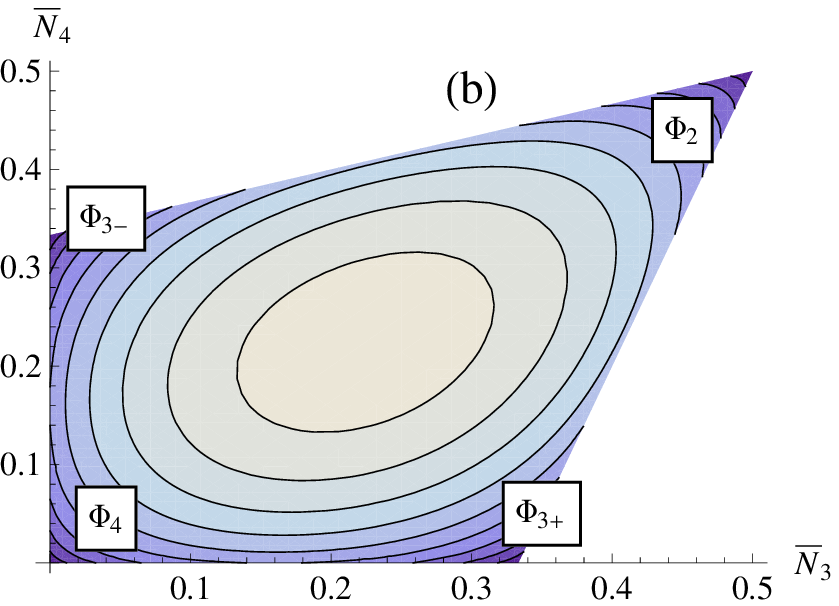}
\caption{(Color online) Entropy per site, $\bar{S}/k_{\mathrm{B}}$, versus population densities,
$\bar{N}_m, \bar{N}_{m'}$, of two species from table~\ref{tab:specsP2}.
The population densities of the other species are kept at $10^{-5}$.
The contours are at (a) $0.070\ell$, (b) $0.071\ell$, $\ell=1,\ldots,6$.}
  \label{fig:picnnn-2-enta}
\end{figure}

All four species of compacts from section~\ref{sec:nnnp2} interlink with elements of pseudo-vacuum
but only selectively with themselves or each other.
This produces two features in their entropy landscapes not seen in the previous cases.
Compacts 1 and 2, which interlink with themselves but not with each other or with any other compact,
generate a triangular entropy landscape as shown in figure~\ref{fig:picnnn-2-enta}~(a).
The entropy is nonzero along each leg, where compacts from one species mix with
elements of pseudo-vacuum.
The entropy is zero along the hypotenuse, where compacts are close-packed.
The system has the highest capacity for compacts 1 and 2 if they are segregated.

The most remarkable entropy landscape pertains to compacts 3 and 4 as shown in
figure~\ref{fig:picnnn-2-enta}~(b).
These two compacts interlink with each other but not with themselves or with any other compact.
The border is a quadrilateral as already seen in figure~\ref{fig:picnnn-1-enta}~(a) but with the sides away
from the axes now slanted positively.
If we start filling the system with compacts 3, the entropy first increases from zero
and then decreases again down to zero when capacity is reached at $N_3=\frac{1}{3}N$ in a
uniformly stacked array.
Interestingly, we do not have to remove any compacts 3 to make space for compacts 4.
In fact, for every three compacts 4 added we can add one more compact 3.
The entropy rises in the process and then returns to zero when the numbers of both species
have become equal and reached the value $\frac{1}{2}N$.

\begin{figure}[!b]
\begin{center}
 \includegraphics[width=85mm]{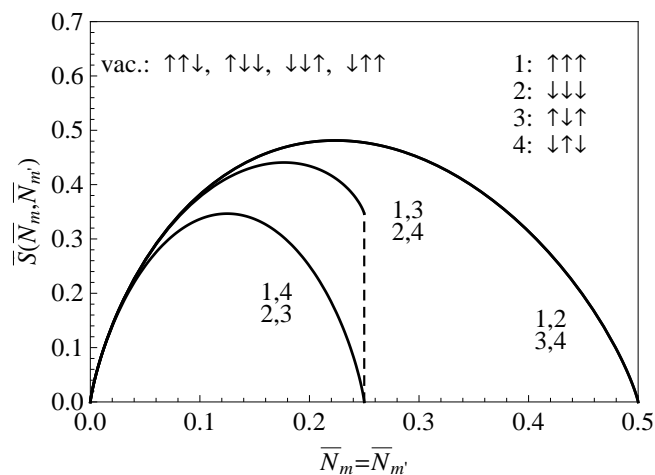}
 \end{center}
\caption{Entropy $\bar{S}/k_{\mathrm{B}}$ per lattice site as function of the density $\bar{N}_m=\bar{N}_{m'}$
of two species of compacts from table~\ref{tab:specsP2}.}
  \label{fig:picnnn-9a}
\end{figure}
Consider a macroscopic system initially in the pseudo-vacuum of compacts.
What happens to the entropy if we add equal numbers of particles
from two of the four species until capacity is reached?
The answer is shown in figure~\ref{fig:picnnn-9a}.
At low density, the positional disorder is little affected by the different shapes of the particles.
The curves overlap close to perfectly.
At higher densities the different shapes dictate the presence or absence of
ordering tendencies and the types of ordering realized.
Particles 1,1 or 2,2 or 3,4 interlink directly.
Two close-packed particles 1,2 or 3,3 or 4,4 have two vacuum elements in a particular
sequence between them.

With increasing (averaged) densities $\bar{N}_1=\bar{N}_2$ or $\bar{N}_3=\bar{N}_4$ the entropy varies
along the same curve even though the associated equilibrium states are very different.
In the case of particles 1,2 the shapes favor single-species clustering when the space becomes crowded.
Mixed-species clustering is favored in the case of particles 3,4.
When capacity is reached, the entropy has returned to zero.
The equilibrium state is then fully phase separated in one case, consisting of two equal-size
single-species clusters, or homogeneous in the other case, consisting of one mixed-species cluster.

Particles 1,4 (or 2,3) are close-packed with three vacuum elements between them.
By contrast, close-packed particles 1,1 interlink directly and close-packed particles 4,4
have two vacuum elements between them.
With increasing (averaged) densities $\bar{N}_1=\bar{N}_4$, single-species clustering crowds out
mixed-species clustering.
Capacity is reached earlier.
The entropy returns to zero as before.
The equilibrium is fully phase-separated.
The two clusters are of unequal size.

Particles 1,3 (or 2,4) are separated by only one vacuum element when close-packed.
The consequence is that with increasing (averaged) densities $\bar{N}_1=\bar{N}_3$ single-species
and mixed-species clustering are equally favorable (compared to loose
particles).
The system reaches capacity in an amorphous state.
The entropy stays nonzero.
The higher rise of the entropy compared to the previous case is explained by the
smaller size of close-packed mixed-species pairs, which increases the positional
disorder at equilibrium.

%
\section{Statistical mechanics}\label{sec:statmech}
%
The statistical mechanical analysis of the particles with shapes identified in section~\ref{sec:motspecat}
can be performed for open or closed systems.
Here we consider an open system.
Wu's analysis \cite{Wu94} for a generic situation starts from the expression
\begin{equation}\label{eq:Zgen}
Z=\sum_{\{N_m\}}W(\{N_m\})\exp\left(-\sum_{m}\frac{\epsilon_mN_m}{k_{\mathrm{B}}T}\right)
\end{equation}
for the grandcanonical partition function, where $\epsilon_m$ are the particle energies
and $W(\{N_m\})$ the multiplicity function.\footnote{The chemical potential is zero for
all species, a consequence of the energy scales
chosen in section~\ref{sec:motspecat}.}
That analysis produces the general result
\begin{equation}\label{eq:3-1}
Z=\prod_{m}\left(\frac{1+w_m}{w_m}\right)^{A_m},
\end{equation}
where the (real, positive) $w_m$ are the solutions of the coupled nonlinear algebraic equations,
\begin{equation}\label{eq:3-2}
\frac{\epsilon_m}{k_{\mathrm{B}}T}=\ln(1+w_m)-\sum_{m'}g_{m'm}\ln\left(\frac{1+w_{m'}}{w_{m'}}\right).
\end{equation}
The capacity constants $A_m$, and the mutual interaction coefficients
$g_{mm'}$  are tabulated in section~\ref{sec:motspecat}.
Arbitrary energies $\epsilon_m$ can be assigned to each particle species.
With the $w_m$ from (\ref{eq:3-2}), the average numbers of particles can be derived from
(\ref{eq:Zgen}) via
\begin{equation}\label{eq:Nmgen}
\langle N_m\rangle =-k_{\mathrm{B}}T\frac{\partial\ln Z}{\partial\epsilon_m}\,,
\end{equation}
which, when carried out using (\ref{eq:3-1}) and (\ref{eq:3-2}), leads to the linear coupled
equations\footnote{The analysis assumes that $N\gg1$ and $\langle N_m\rangle\gg1$.
Any part of a nonzero $A_m$ that is of O(1) is irrelevant. Wu's derivation of (\ref{eq:3-1}) proceeds
via (\ref{eq:3-3}). The calculation of (\ref{eq:3-3}) from (\ref{eq:Nmgen}) via (\ref{eq:3-1})
and (\ref{eq:3-2}) is merely a check of internal consistency.}
\begin{equation}\label{eq:3-3}
w_m\langle N_m\rangle+\sum_{m'}g_{mm'}\langle N_{m'}\rangle =A_m\,.
\end{equation}
In similar fashion we can derive from (\ref{eq:3-1}) and (\ref{eq:3-2}) correlations between particle
populations, specifically the covariances
\begin{equation}\label{eq:covariance}
\langle\langle N_mN_{m'}\rangle\rangle\doteq
\langle N_mN_{m'}\rangle -\langle N_m\rangle\langle N_{m'}\rangle=(k_{\mathrm{B}}T)^2
\frac{\partial^2\ln Z}{\partial\epsilon_m\partial\epsilon_{m'}}\,.
\end{equation}
The entropy inferred from (\ref{eq:3-1}) can be expressed as a function of the
$\langle N_{m}\rangle$ alone, namely by the function $S(\{\langle N_{m}\rangle\})$ from
(\ref{eq:snmp}).

It is instructive to compare this method with the transfer matrix method \cite{Yeom92}
in the context of the Ising chain (\ref{eq:1}).
The latter operates with coupled degrees of freedom of minimal structure whereas
the former operates with degrees of freedom that are no longer coupled but have more complex
structures.
Ising spins are tied to lattice sites whereas particles are floating.
The number $N$ of Ising spins is fixed whereas the numbers $N_m$ of particles from each species
are fluctuating.
Ising spins at different sites are distinguishable whereas particles from the same species are not.
The canonical partition function $Z_N$ from the transfer matrix analysis is related to the
grand partition function (\ref{eq:3-1}) via
\begin{equation}\label{eq:Zequiv}
Z=\re^{E_{\mathrm{pv}}/k_{\mathrm{B}}T}Z_N\,,
\end{equation}
where $E_{\mathrm{pv}}$ is the energy of the pseudo-vacuum in use.

For a brief demonstration of how particles with interlinking motifs and different energies compete
for space in the presence of thermal fluctuations and produce long-range ordering tendencies
at low $T$ as a consequence, we assign the energies of (\ref{eq:1}) at $h=0$ as tabulated.
All thermodynamic quantities of interest can then conveniently be expressed in terms of the
function\footnote{The mapping which relates the
special cases $h=0$ and $L=0$ of Hamiltonian (\ref{eq:1}) with open
boundary conditions by simple parameter transcription $J\leftrightarrow-h$, $L\leftrightarrow J$
\cite{dobsteph} is evident in the partition function but has no impact on what is demonstrated here.}
\begin{equation}\label{eq:ukjkl}
u(K_J,K_L)=\cosh K_J+\sqrt{\sinh^2K_J+\re^{4K_L}}\,,\qquad
K_J\doteq\frac{J}{4k_{\mathrm{B}}T}\,,\qquad K_L\doteq\frac{L}{4k_{\mathrm{B}}T}\,.
\end{equation}
The familiar transfer matrix result then reads  \cite{Yeom92}
\begin{equation}\label{eq:tmzn}
Z_N=\left[u(K_J,K_L)\re^{-K_L}\right]^N.
\end{equation}

\subsection{Analysis from
$|\uparrow\downarrow\uparrow\downarrow\cdots\rangle_2$}\label{sec:nnnp1sm}
The physically relevant solution of equations~(\ref{eq:3-2}) for the four species of hosts and tags
introduced in section~\ref{sec:nnnp1} is
\begin{align}\label{eq:213-3}
w_1=w_2=u(K_J,K_L)\re^{K_J}-1,\qquad w_3=w_4=w_1\re^{-4K_L}.
\end{align}
The grand partition function (\ref{eq:3-1}) becomes
\begin{align}\label{eq:213-4}
Z=\left[u(K_J,K_L)\re^{-K_J}\right]^N,
\end{align}
consistent with (\ref{eq:tmzn}) via (\ref{eq:Zequiv}).
Equations~(\ref{eq:3-3}) yield
\begin{align}\label{eq:214-1}
\langle\bar{N}_1\rangle =\langle\bar{N}_2\rangle =\langle \bar{N}_3\rangle\frac{1}{w_1}\,, \qquad
\langle \bar{N}_3\rangle =\langle \bar{N}_4\rangle =\frac{1}{2}\frac{w_1}{w_1w_3+2w_1+1}\,.
\end{align}

\begin{figure}[htb]
  \includegraphics[width=68mm]{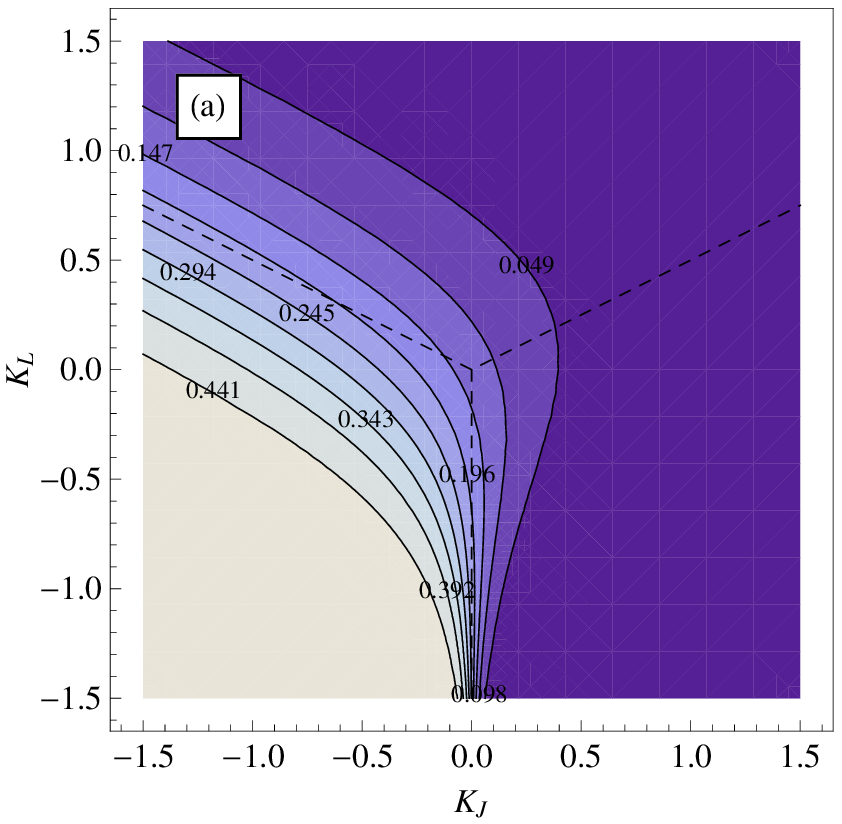}\hspace{5mm}%
  \includegraphics[width=68mm]{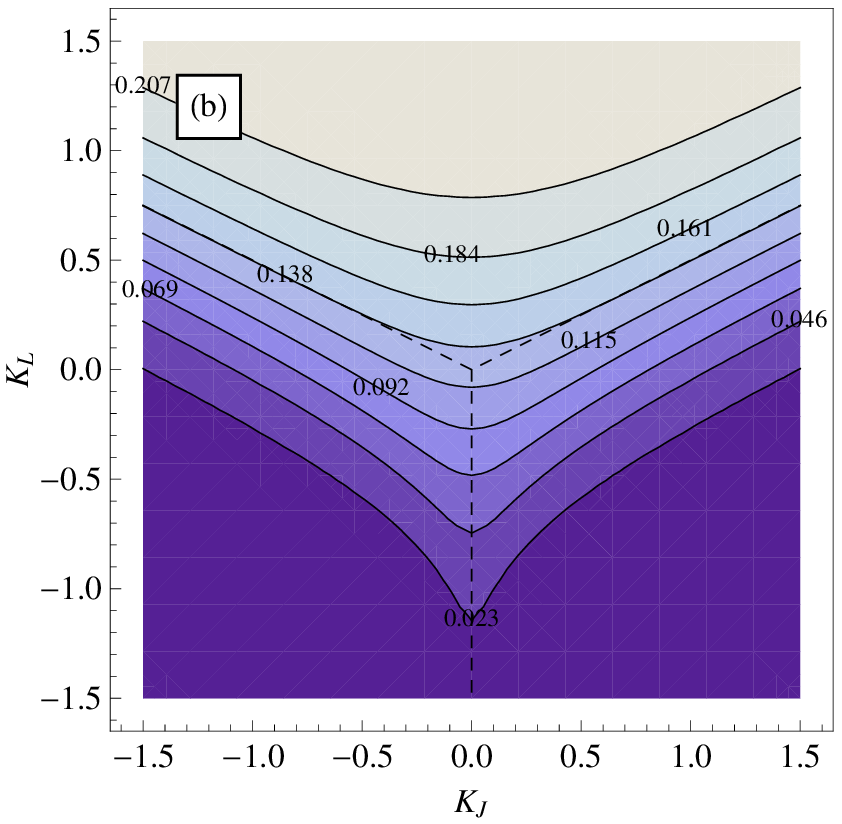}
\caption{(Color online) Average number (per site) (a) $\langle \bar{N}_{1}\rangle= \langle
  \bar{N}_{2}\rangle$ of tags and (b) $\langle\bar{N}_{3}\rangle=\langle \bar{N}_{4}\rangle$ of hosts
  versus $K_J, K_L$. Plot (a) also represents the population density of FM particles discussed in
   section~\ref{sec:nnnp2sm}. The population density of AFM particles from that same set is the left-right
   mirror image of plot (a).  The dashed lines indicate the $T=0$ phase boundaries of
  figure~\ref{fig:plj5f1c}~(b).}
  \label{fig:plj5-2_1_5}
\end{figure}

In figure~\ref{fig:plj5-2_1_5} we show contour plots of the population densities of tags
$m=1,2$ $(\uparrow\uparrow\uparrow,\downarrow\downarrow\downarrow)$ and
hosts $m=3, 4$ $(\downarrow\uparrow\uparrow\downarrow,\uparrow\downarrow\downarrow\uparrow)$.
Lowering $T$ at fixed $J,L$ means moving from the center radially outward in a given direction.
Any particular population density either increases or decreases from the
common value $\langle \bar{N}_m\rangle=\frac{1}{8}$ at $T=\infty$.
In region $\Phi_2$ all $\langle \bar{N}_m\rangle$ decrease and reach zero at $T=0$.
All particle energies $\epsilon_m$ are positive.
Throughout region $\Phi_4$ the host energies are negative.
The host population densities $\langle \bar{N}_3\rangle=\langle \bar{N}_4\rangle$
increase as $T$ is lowered and reach the value $\frac{1}{4}$ at $T=0$.
In part of this region, at $J<0$, the tags have negative energies as well, but their
energy density is less negative than that of hosts.
Hence tags are crowded out by hosts.
In region $\Phi_1$ the host population diminishes and the tag population proliferates
toward $\langle\bar{N}_1\rangle=\langle\bar{N}_2\rangle=\frac{1}{2}$ as $T$ is lowered.
In part of this region, for $L>0$, the hosts have lower (negative) excitation energies than the tags.
Nevertheless, the hosts are crowded out by the tags due to their larger size.

On the $\Phi_1-\Phi_2$ boundary ($J=0, L<0$) the hosts have $\epsilon_m>0$ and the tags
$\epsilon_m=0$.
As $T$ is lowered the host population decreases, leaving more room for tags and elements of
pseudo-vacuum, whose motifs do not interlink directly.
At low $T$ the few remaining hosts act as surfactants between segments of tags inside and
segments of pseudo-vacuum outside \cite{copic}.
The ground state is fourfold degenerate. Two vectors contain no particles.
The other two are solids formed by tags of one or the other kind.
The density of tags averaged over the four vectors is $\langle \bar{N}_1\rangle =\langle
\bar{N}_2\rangle=\frac{1}{4}\,$.

On the $\Phi_2-\Phi_4$ boundary ($L=J/2>0$) the particles with $\epsilon_m>0$ are tags and the
particles with $\epsilon_m=0$ are hosts -- a switch with drastic consequences.
As $T$ is lowered the tag population decreases, surrendering the lattice to hosts and elements
of pseudo-vacuum, whose motifs do interlink directly.
A highly degenerate ground state ensues.
The limiting host population densities (\ref{eq:214-1}) are
$\langle\bar{N}_3\rangle =\langle\bar{N}_4\rangle =(5+\sqrt{5})^{-1}\simeq 0.138$.

Finally, on the $\Phi_1-\Phi_4$ boundary ($L=-J/2>0$), all particles have $\epsilon_m<0$.
Hosts and tags have equal negative energy densities.
All configurations of hosts and tags covering the lattice solidly have the same
energy.
The resultant entropy is the same as that along the $\Phi_2-\Phi_4$ boundary.
The host population density remains the same,
$\langle\bar{N}_3\rangle =\langle\bar{N}_4\rangle =(5+\sqrt{5})^{-1}\simeq 0.138$.
The tag population density, calculated from (\ref{eq:214-1}), is
$\langle\bar{N}_1\rangle =\langle\bar{N}_2\rangle =(2\sqrt{5})^{-1} \simeq 0.224$.

\subsection{Analysis from
$|\uparrow\uparrow\downarrow\downarrow\uparrow\uparrow\cdots\rangle_4$}\label{sec:nnnp2sm}
The grand partition function for the four compacts introduced in section~\ref{sec:nnnp2}
becomes
\begin{align}\label{eq:223-4}
Z =\left[u(K_J,K_L)\re^{-2K_L}\right]^N,
\end{align}
again consistent with (\ref{eq:tmzn}) via (\ref{eq:Zequiv}), and where
\begin{align}\label{eq:223-3}
w_1=w_2=u(K_J,K_L)\re^{K_J}-1,\qquad w_3=w_4=(1+w_1)\re^{-2K_J}-1,
\end{align}
are the solutions of  equations~(\ref{eq:3-2}), and equations~(\ref{eq:3-3}) yield
\begin{align}\label{eq:224-1}
\langle \bar{N}_1\rangle =\langle \bar{N}_2\rangle =\frac{1}{2}\frac{w_3}{2w_1w_3+w_1+w_3}\,,\qquad
\langle \bar{N}_3\rangle =\langle \bar{N}_4\rangle =\frac{1}{2}\frac{w_1}{2w_1w_3+w_1+w_3}\,.
\end{align}

The population densities of the  FM particles 1,2 $({\uparrow\uparrow\uparrow,
\downarrow\downarrow\downarrow})$ and the AFM particles 3,4
$({\uparrow\downarrow\uparrow, \downarrow\uparrow\downarrow})$ have a mirror-image
relationship (figure~\ref{fig:plj5-2_1_5}).\footnote{Tag 2 from table~\ref{tab:specsP1} and FM
compact 2 from table~\ref{tab:specsP2} have identical motifs and, therefore, identical
(absolute) energies. However, they are different particles existing in
different worlds. Nevertheless, their population densities are identical at $h=0$.}
Lowering $T$ from infinity either leads to a depletion of FM particles (region
$\Phi_2$) or to a depletion of AFM particles (region $\Phi_1$) or to a depletion
of both kinds of particles (region $\Phi_4$).

If $T$ is lowered on the $\Phi_1-\Phi_2$ border, all four particles have equal and
negative energies.
What matters for the nature of the ground state is that FM particles and AFM particles
interlink among themselves (albeit in different manner) but not with each other.
The ground state thus consists of the four states that are either packed with FM
particles or with AFM particles.

Along the $\Phi_1-\Phi_4$ border, FM particles have zero energy and AFM
particles have positive energies.
The lowest energy level is highly degenerate, comprising states populated by FM
particles in all permissible configurations.
The average population density of these compacts is
$\langle\bar{N}_1\rangle =\langle\bar{N}_2\rangle =(2\sqrt{5})^{-1} \simeq 0.224$.
The role of the AFM particles along the $\Phi_2-\Phi_4$ border is the same as that of the
FM particles along the $\Phi_1-\Phi_4$ border in all respects
except the way the motifs interlink.

\subsection{Analysis from $|\uparrow\uparrow\cdots\rangle_1$}\label{sec:nnnp0sm}
Equations (\ref{eq:3-2}) of the five species of hosts, tags, and hybrids introduced in
section~\ref{sec:nnnp0} have the solutions
\begin{subequations}\label{eq:233-6}
\begin{align}
&w_3 =u(K_J,K_L)\re^{K_J}-1,\qquad
\frac{1+w_4}{(1+w_3)^2}=\re^{-4K_J},\qquad \frac{1+w_5}{w_3w_4}=\re^{-4K_L}, \\
& w_1=\frac{w_4w_5}{1+w_4}\re^{-4K_L},\qquad w_2=\frac{1+w_1}{w_1w_3-1}\,,
\end{align}
\end{subequations}
implying
\begin{equation}\label{eq:233-3}
Z=\left[u(K_J,K_L)\re^{K_J}\right]^N.
\end{equation}

The solutions of equations~(\ref{eq:3-3}) are again straightforward but, in this instance, unwieldy.
Symmetries dictate that $\langle\bar{N}_2\rangle=\langle\bar{N}_5\rangle$.
The population densities as presented graphically in figure~\ref{fig:plj5-2_3_5}  are more complex.
Two of the species play only minor parts in the statistical mechanics.
Hosts 2 $(\uparrow\uparrow\downarrow\uparrow\uparrow)$ and hybrids 5
$(\downarrow\downarrow\uparrow\downarrow\downarrow)$ are not abundantly present
for any combination of coupling constants or any temperature.
They barely make it to an average of five particles per hundred lattice sites under the most
favorable equilibrium circumstances, whereas hosts 1
$(\uparrow\uparrow\downarrow\downarrow\uparrow\uparrow)$ make it to an average of
up to 25 and the tags 3 $(\downarrow\downarrow\downarrow)$, 4
$(\uparrow\downarrow\uparrow\downarrow, \downarrow\uparrow\downarrow\uparrow)$
 to averages of up to 50.

At $L>|J|/2$, host 1 has the lowest energy of all by a considerable margin.
Its population density is dominant in that region.
At $J>0, L<0$ the equilibrium state is dominated by tags 4 and at $J<0, L<0$ by tags 3.
In both regions these low-energy tags can only exist as attachments to higher-energy hosts.
As $T$ is lowered fewer and fewer hosts have more and more tags attached to them.
At $T=0$ the last host particle is replaced by a low-energy tag in a configuration where each
tag is attached to its neighbor on the left around the chain.

In the remaining two regions at $0<L<|J|/2$ two or three species of particles are in competition
for dominance with energy, size, and category all factoring in.
The smallest size among the low-energy particles happens to win in both regions as the
temperature is turned down: tag 3 at $J<0$ and tag 4 at $J>0$.
There is no region where the total particle population is thinned out entirely as $T\to0$.
Tags 3 have zero energy.
They only disappear if they are crowded out by particles with negative energy.
Our choice of a pseudo-vacuum that does not coincide with the physical vacuum.

Along the $\Phi_1-\Phi_2$ border tags 3,4 both have zero energy while all
other particles have positive energies.
Tags 3 do not interlink with tags 4.
The ground-state degeneracy remains low (fourfold).
On the $\Phi_1-\Phi_4$ border we have again two zero-energy species (hosts 1 and tags 3) and three
positive-energy species.
However, the former two do interlink.
The ground-state degeneracy is high.
The population densities are $\langle\bar{N}_1\rangle=(5+\sqrt{5})^{-1}\simeq 0.138$
for the hosts and $\langle\bar{N}_3\rangle= (2\sqrt{5})^{-1}\simeq 0.224$ for the tags.

A considerably more complex scenario unfolds on the $\Phi_2-\Phi_4$ border.
Four species of particles of three different sizes and with three different negative
energies are in competition.
Only tag 3 (with zero energy) stands by.
Of the four competing particles, 1,2,5 can host and  4,5 can be hosted.
Hosts 2 and hybrids 5 are absent at $T=0$ except on the phase boundary.
There we have $\langle\bar{N}_1\rangle=\langle\bar{N}_4\rangle=\frac{1}{9}\,$,
$\langle\bar{N}_2\rangle=\langle\bar{N}_5\rangle=\frac{1}{18}\,$.
At $T>0$ the particles 2,5 maintain significant populations only near the
phase boundary.

\subsection{Entropy}\label{sec:entag}
\begin{figure}[htb]
  \includegraphics[width=68mm]{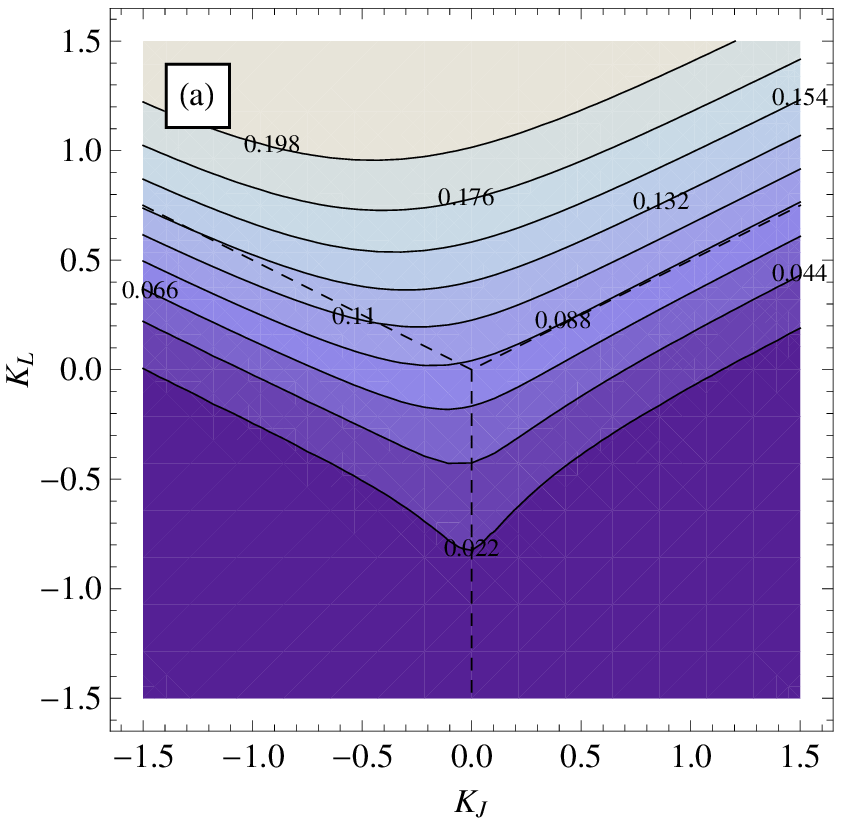}\hspace{3mm}%
  \includegraphics[width=68mm]{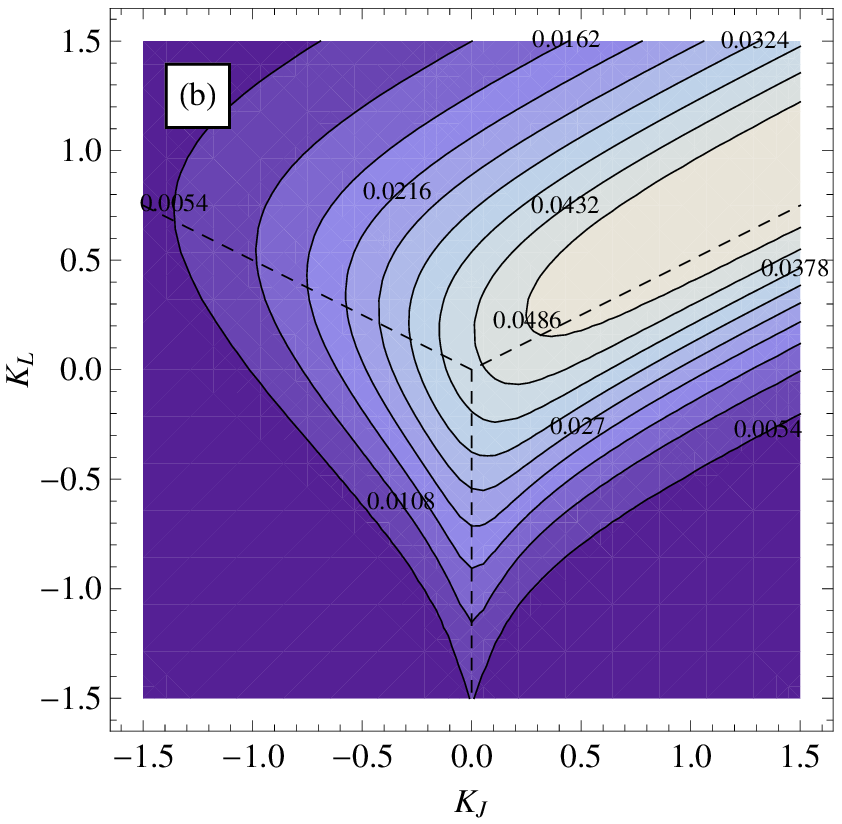}
  \includegraphics[width=68mm]{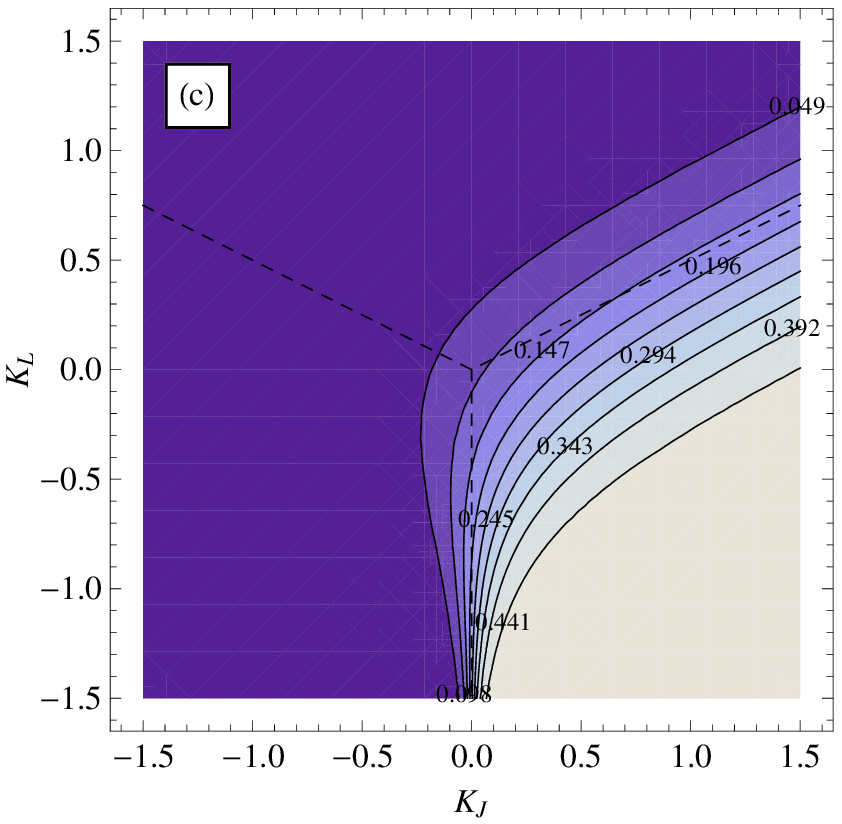}\hspace{3mm}%
  \includegraphics[width=68mm]{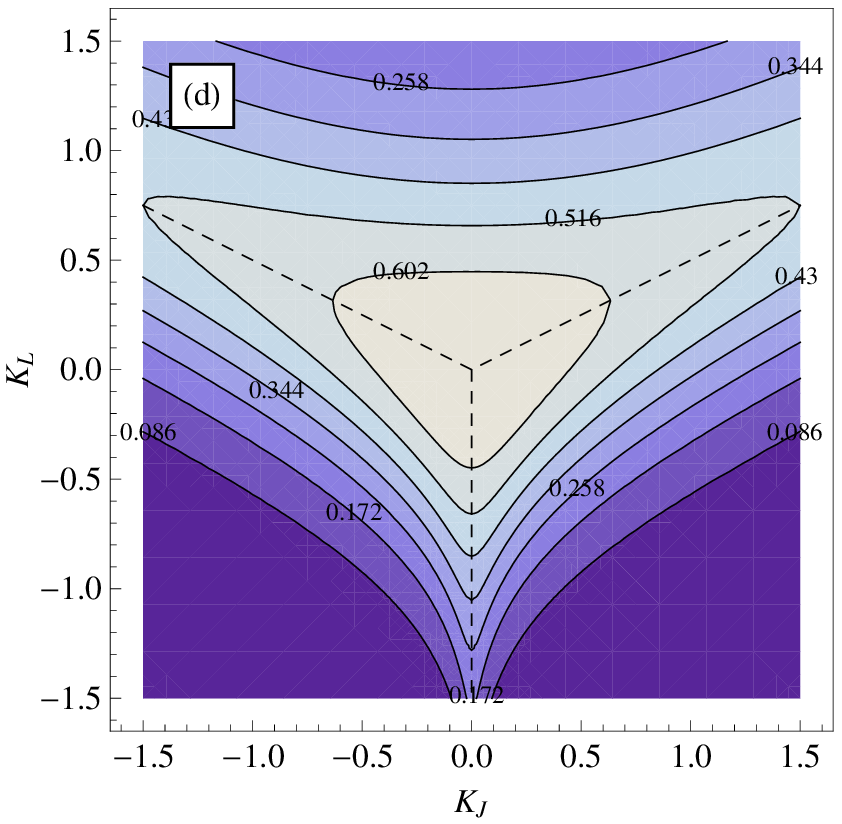}
\caption{(Color online) Average number (per site) (a) $\langle \bar{N}_{1}\rangle$,
(b) $\langle\bar{N}_{2}\rangle=\langle \bar{N}_{5}\rangle$, and (c)  $\langle \bar{N}_{4}\rangle$
  versus scaled variables $K_J, K_L$. The corresponding plot for $\langle \bar{N}_{3}\rangle$
  is as shown in figure~\ref{fig:plj5-2_1_5}~(a). The graphs of $\langle \bar{N}_{3}\rangle$ and
  $\langle \bar{N}_{4}\rangle$ are not mirror images of each other. The graph of
  $\langle \bar{N}_{1}\rangle+\langle \bar{N}_{2}\rangle$ is identical to the graph shown in
  figure~\ref{fig:plj5-2_1_5}~(b).
  Panel (d) shows the entropy $\bar{S}/k_{\mathrm{B}}$ per lattice site as function of the scaled variables
  $K_J, K_L\,$.}
  \label{fig:plj5-2_3_5}
\end{figure}

The three sets of particles produce distinct entropy landscapes in the space of population densities
as discussed in section~\ref{sec:entlan}.
This diversity is attributable to the different structures of the particle species and their mutual
statistical interactions as encapsulated in the entropy expression (\ref{eq:snmp}).
However, when applied to a particular physical context, where all particle energies $\epsilon_m$
are functions of Hamiltonian parameters, the function $S(\{\langle N_m\rangle\})$ inferred
from (\ref{eq:snmp})\footnote{Alternatively derived from (\ref{eq:tmzn}) via $S=k_{\mathrm{B}}\rd T\ln Z_N/\rd T$.}
 encodes, at any given temperature, the same entropy landscape for all three
sets of particles in the space of these parameters.
That landscape is shown in figure~\ref{fig:plj5-2_3_5}~(d) for the situation at
hand.

For given $J$ and $L$ the entropy is a smooth and monotonously increasing
function of $T$.
Magnetic short-range order (one of three kinds) establishes itself gradually and turns into
long-range order at $T=0$.
The thermal fluctuations are stronger at $L>0$ than at $L<0$ if $J\neq0$ owing to
competing \textsf{nn} and \textsf{nnn} couplings.
$\bar{S}/k_{\mathrm{B}}$ decreasing smoothly from $\ln2\simeq 0.693$ at ${T=\infty}$
to zero at $T=0$ for all parameter combinations except $|J|=2L$.
Here $\bar{S}/k_{\mathrm{B}}$  approaches $\ln([1+\sqrt{5}]/2)\simeq 0.481$ in the low-$T$ limit.
The implied ground-state degeneracy has a natural yet different interpretation in the context of each set
of statistically interacting particles as shown.

%
\section{Structures from interactions}\label{sec:strucint}
%
In this work we have investigated structures caused by interactions in different ways.
On a small scale, particles of various shapes are assembled from building blocks with
minimal structure (Ising spins) by the \textsf{nn} and \textsf{nnn} couplings of Hamiltonian (\ref{eq:1}).
These particles with structures over the range of a few lattice sites are free of any
inter-particle binding forces, not merely at low density but also at high density.
Nevertheless, by virtue of their specific shapes, these particles are apt to assemble further
structures on intermediate and large scales in the form of positional ordering of various kinds.

On an intermediate scale we have described the assembly of clusters of compacts,
a process driven entirely by shapes and limited space.
We have also described host particles consisting of two amphiphilic parts bracketing
uniform arrays of tags or disordered mixtures of tags and hybrids, thus assuming the role
of a surfactant or a membrane.
Host particles are akin to micelles in this context.
On a larger scale we have described the formation of macroscopically ordered patterns
of particles from one or two species in a crowded environment.
The type of ordering is determined solely by the energies, shapes, and sizes of the particles.

The structures stabilized on the smallest scale by the coupled Ising spins depend on our
choice of reference state (pseudo-vacuum), hence the three sets of particle species in
tables~\ref{tab:specsP1}, \ref{tab:specsP0},
\ref{tab:specsP2}.
However, in any given equilibrium state, the same large-scale ordering tendencies are produced
by particles from different sets, i.e. by particles with different shapes, sizes, and energies.

The methodology developed here and in reference~\cite{copic} has natural applications in research
areas of strong current interest including jamming of granular matter in narrow channels~\cite{AB09} and DNA overstretching~\cite{VMRW05,sarkar}.



\ukrainianpart

\title{Взаємозв'язані мотиви і ландшафти ентропії статистично взаємодіючих
частинок}

\author{П. Лу\refaddr{label1}, Д. Ліу\refaddr{label1},
       Г. Мюллер\refaddr{label1}, M. Карбах\refaddr{label2}}

\addresses{
\addr{label1} Фізичний факультет, Університет Род-Айленду, Кінгстон Род-Айленд
02881, США
\addr{label2} Фізичний факультет, Гірничий університет Вупперталя, 42097
Вупперталь, Німеччина}

\makeukrtitle

\begin{abstract}
\tolerance=3000%
Використовуючи $s = 1/2$ iзингiвський  ланцюжок з однорiдною взаємодiєю найближчих i наступних за найближчими  сусiдiв, побудовано систему незафiксованих частинок, якi характеризуються мотивами шести послідовних локальних спiнiв. Взаємодiя спiнiв спричиняє групування частинок, якi, в свою чергу, не залежать вiд енергiй взаємодiї навiть при високiй густинi. Всi  мiкростани утворені конфiгурацiями  частинок з одного iз трьох рiзних  наборiв, які відповідають збудженням псевдовакуумів, що пов'язані iз основними станами перiодичностi один, два i чотири.  Мотиви частинок i елементи псевдовакууму об'єд\-нуються  в двох спiльних вузлових змiнних. Статистична взаємодiя мiж частинками є закодованою в узагальненому  принципi Паулi, що описує як розмiщення  одної частинки змінює можливості для розмiщення подальших частинок. В статистично механiчному  аналiзi довiльнi енергiї можуть ставитись у вiдповiднiсть всiм сортам частинок. Ентропiя є функцiєю заселеностi частинок. Особливостi статистичної взаємодiї прозоро вбудованi в цей вираз. Енергiї i структури частинок виключно визначають впорядкування при низьких температурах. За особливих умов частинки можуть бути замiненi фундаментальнiшими  частинками з коротшими мотивами, що взаємозв'язуються однією спiльною вузловою змiнною. Зумовлені взаємодією структури виникають на двох рiвнях:  частинки з формами, які утворені взаємодіючими спiнами, i тенденцiї до далекосяжного впорядкування частинок з формами, які статистично взаємодiють.
\keywords принцип Паулі, частинки з формами, дробова статистика, модель Ізинга, солітони

\end{abstract}
\end{document}